\documentstyle[12pt,epsfig]{cernart}
\makeatletter
\def\thebibliography#1{\section*{\refname\@mkboth
 {\uppercase{\refname}}{\uppercase{\refname}}}\list
 {\@biblabel{\arabic{enumi}}}{\settowidth\labelwidth{\@biblabel{#1}}%
 \leftmargin\labelwidth
 \advance\leftmargin\labelsep
 \usecounter{enumi}% \let\p@enumi\@empty
 \def\theenumi{\arabic{enumi}}}%
 \def\newblock{\hskip .11em plus.33em minus.07em}%
 \sloppy\clubpenalty4000\widowpenalty4000
 \sfcode`\.=1000\relax}
\makeatother
\def\Aref#1{$^{\rm #1}$} 
\def\AAref#1#2{$^{\rm #1,#2}$} 
\def\IAref#1#2{$^{\Inst{#1},\rm #2}$} 
\def\IIref#1#2{$^{\Inst{#1},\Inst{#2}}$} 

\def\IIAref#1#2#3{$^{\Inst{#1},\Inst{#2},\rm #3}$}
\def\Iref#1{$^{\Inst{#1}}$} 
\begin {document}
\begin {titlepage}
\docnum {CERN--PPE/94--116}
\date {14 July 1994}
\vglue2cm
\title{SPIN ASYMMETRY IN MUON--PROTON DEEP INELASTIC SCATTERING
ON A TRANSVERSELY-POLARIZED TARGET}
\vglue2cm
\collaboration {The Spin Muon Collaboration (SMC)}

\vglue4cm
\begin{abstract}
We measured the spin asymmetry in the
scattering of 100 GeV longitudinally-polarized muons on
transversely polarized protons.
The  asymmetry  was   found to be
compatible with zero in the kinematic range
$0.006<x<0.6$,  $1<Q^2<30\,~\mbox{GeV}^2$.
{}From this result we derive the
upper limits for the virtual photon--proton asymmetry $A_2$,
and for the spin structure function
$g_2$. For $x<0.15$, $A_2$ is  significantly
smaller than its positivity limit
$\sqrt{R}$.
\end{abstract}
\vglue 5cm
\submitted {Submitted to Physics Letters B}
\newpage
\vglue 5cm
\input{authors}
%
%\vglue 3cm
\end {titlepage}

The nucleon spin-dependent structure functions, $g_1$ and $g_2$, can be
determined in deep inelastic scattering of polarized charged leptons on
polarized nucleons.  Both are important for the understanding of
nucleon spin structure.  The structure function $g_1$ is used to test
QCD sum rules, and in the quark--parton model it determines the
contribution of the quark spins to the nucleon spin.  The spin
structure function $g_2$ differs from zero because of the masses and the
transverse momenta of the quarks.  It has a unique leading-order
sensitivity to twist-3 operators, i.e., quark--gluon correlation effects
in QCD. The measurement of $g_2$ would provide the first information on
these twist-3 operator matrix elements~\cite{JA}.

In general, $g_2$ can be written as the sum of a contribution, $g_{2}^{\rm
 ww}$,
directly calculable from $g_1$~\cite{WW}, and a pure twist-3 term
$\overline{g_2}$~\cite{JA}
\begin{equation}
    g_2(x,Q^2) = g_{2}^{\rm ww}(x,Q^2) +\; \overline{g_2}(x,Q^2)\;,
\label{g2}
\end{equation}
with
\begin{equation}
    g_{2}^{\rm ww}(x,Q^2) = - g_1(x,Q^2) +\int_{x}^{1} g_1(t,Q^2) \frac{{\rm
 d}t}{t}\;,
\label{g2ww}
\end{equation}
where $-Q^2$ is the four-momentum transfer squared, and $x$ is the Bjorken
 scaling variable.In several recent papers, values for $\overline{g_2}$
and their approximate $Q^2$ dependence have been calculated using different
assumptions \cite{STRAT,CHJI,ALI,JAJI}.
A sum rule for $g_2$,
\begin{equation}
    \int_{0}^{1} g_2(x,Q^2) {\rm d}x = 0\;,
\label{int0}
\end{equation}
was derived by Burkhardt and Cottingham using Regge theory~\cite{BUCO}.
It has been regarded as a consequence of conservation of angular
 momentum~\cite{Feynman}.
At present the validity of the derivation of this sum rule is in
 question~\cite{MS,JA},
and it is clearly important to test it experimentally.

Two kinds of spin-dependent cross-section asymmetries can be measured in
 inclusive lepton--nucleon
scattering. When the target polarization is parallel to the direction of
the longitudinally-polarized beam, the asymmetry is given by $A_\parallel$,
whilst for a target polarized in a direction transverse to the beam,
the asymmetry is given by $A_T$
\begin{equation}
  A_\parallel\, (x,Q^2)=  \frac
           {{\rm d}\sigma^{\uparrow\downarrow}-{\rm
 d}\sigma^{\uparrow\uparrow}}
           {{\rm d}\sigma^{\uparrow\downarrow}+{\rm
 d}\sigma^{\uparrow\uparrow}}\;,
\hspace{2cm}
  A_T (x,Q^2,\phi) =
        \frac{ {\rm d}\sigma^{\downarrow\rightarrow} -
               {\rm d}\sigma^{\uparrow\rightarrow} }
             { {\rm d}\sigma^{\downarrow\rightarrow} +
               {\rm d}\sigma^{\uparrow\rightarrow} }\;.
\label{AP}
\end{equation}
Here, ${\rm d}\sigma^{\uparrow\uparrow}({\rm d}\sigma^{\uparrow\downarrow})$
 corresponds to ${\rm d}^2\sigma/{\rm d}x\ {\rm d}Q^2$ for
parallel~(antiparallel) beam and target polarization.
For perpendicular target polarisation,
${\rm d}\sigma^{\uparrow\rightarrow}$
(${\rm d}\sigma^{\downarrow\rightarrow}$)
stands for
${\rm d}^3\sigma/{\rm d}x\ {\rm d}Q^2\ {\rm d}\phi$,
when the target spin points in the upwards (downwards) direction.
The azimuthal angle $\phi$ is defined  about the beam axis,
$\phi=0$ corresponding to a muon scattered upwards.
It can be shown that $A_T$ is proportional to $\cos\phi$, and thus
it is convenient to define $A_\perp(x,Q^2) \equiv A_T(x,Q^2,\phi) /  \cos\phi$,
which is independent of $\phi$~\cite{JA}.

The  virtual photon absorption asymmetries
\begin{eqnarray}
  A_1 = \frac{\sigma_{1/2}- \sigma_{3/2}}{\sigma_{1/2}+\sigma_{3/2}}\;,
 \hspace{2cm}
  A_2 = \frac{2\sigma_{TL}}{\sigma_{1/2}+\sigma_{3/2}}\;,
\label{A1}
\end{eqnarray}
are related to the measured asymmetries $A_\parallel$ and $A_\perp$,
\begin{eqnarray}
\label{AG1}
    A_\parallel &=& D \left(A_1+\gamma\frac{(1-y)}{1-y/2}A_2\right) \\
\label{AG2}
    A_\perp &=& \, d \ \left(A_2-\gamma\,(1-\frac{y}{2})\,A_1\right)\;.
\end{eqnarray}
Here $\sigma_{1/2}$ and $\sigma_{3/2}$ are the virtual photon--nucleon
 absorption
 cross sections
for total helicity 1/2 and 3/2, respectively, and $\sigma_{TL}$
arises from the helicity spin-flip amplitude in forward photon--nucleon
Compton scattering~\cite{IOFFE,CLOSE}.
The kinematic factor $\gamma$ is defined by
$\gamma=\sqrt{Q^2}/{\nu}=2Mx/\sqrt{Q^2}$,
where $\nu$ is
the energy transfer in the laboratory frame,
and ~$y=\nu/E_{\mu}$.
The coefficient $d$ is related to the virtual photon depolarization
factor $D$ by
\begin{eqnarray}
d = D\frac{\sqrt{1-y}}{1-y/2} \hspace{1cm} D=\frac{ y(2 - y)}{ y^2 + 2 (1-y)(1
 +
 R)}\;,
\label{d}
\end{eqnarray}
where $R$ is the ratio of the longitudinal to transverse photoabsorption cross
sections, $\sigma_L$/$\sigma_T$.
In the case of $A_\parallel$, the contribution of the
$A_2$ term relative to $A_1$ is suppressed by a factor~$\gamma$.  The opposite
 is
true for the case of $A_\perp$.

Positivity conditions~\cite{POS} limit the magnitudes of $A_1$ and $A_2$
\begin{equation}
    |A_1|<1\;, \qquad |A_2| \leq \sqrt{R}\; .
\label{UPPER}
\end{equation}
The asymmetries $A_1$ and $A_2$ can be expressed in terms of the structure
 functions  $g_1$ and $g_2$
\begin{eqnarray}
    A_1 = \frac{1}{F_1}(g_1-\gamma^2g_2)\;, \hspace{1cm}
    A_2 = \frac{\gamma}{F_1}(g_1+g_2)\;,
\label{A2}
\end{eqnarray}
where $F_1=F_2(1+\gamma^2)/2x(1+R)$ is the spin-independent
structure function.
The ratio $R$, determined at SLAC~\cite{WHIT}, is about 0.3
in the range $0.1<x<0.4$ for $Q^2\cong1.5$~GeV$^2$,
and similar values are usually assumed at smaller~$x$.
Using these values of $R$, the positivity condition on
 $A_2$~[Eq.~(\ref{UPPER})]
combined with Eq.~(\ref{A2}) leads to an upper limit for $|g_2|$ which is
approximately of the form $|x^2g_2| < K$, where $K$ varies from 0.07 to 0.10.
Thus, large values of $g_2$ are allowed in the low $x$ region.

In the past, only $A_\parallel$ has been measured in polarized deep inelastic
experiments on the proton.  The asymmetry $A_1$ has been extracted from these
measurements by neglecting the contribution of the $A_2$ term.
In this paper we present data on $A_\perp$, which makes it possible to extract
$A_1$ and $A_2$
with no approximations [Eqs.~(\ref{AG1}) and~(\ref{AG2})]. The $A_1$ results
have been published in Ref.~\cite{SMC94}. In this Letter, we present the
results for the asymmetry $A_2$, and for the structure function $g_2$.

The experiment was carried out at the CERN SPS by
scattering 100~GeV longitudinally-polarized muons off
transversely-polarized protons.
The polarized target and the spectrometer used for these measurements are
basically the same as those used to measure longitudinal asymmetries,
and have been described in a previous
publication~\cite{SMC94}.
The new SMC target incorporates a sufficiently strong dipole field of
 0.5\,T~\cite{SACLAY}, in which the proton polarization can be maintained
 transverse to the beam direction.
The two cells of the butanol target, each 60\,cm long, were longitudinally
 polarized along a solenoid field
of 2.5\,T by dynamic nuclear polarization (DNP).
When  high polarization was reached, the
proton spins were `frozen' at a base temperature of about 60\,mK, and rotated
to a transverse (vertical) direction by applying the additional
dipole field and reducing the longitudinal field to zero.
Upwards (downwards) transverse polarization was achieved
by choosing the initial longitudinal polarization parallel (antiparallel)
to the beam direction.
The spins were reversed 10 times during the 17 days of
data-taking.
The transverse
field was always applied in the same direction
to avoid different acceptances.
As it was not possible to measure the  polarization in the  transverse
spin direction at 0.5~T, it was measured before and after
each reversal in the solenoid field at 2.5~T.
The loss of polarization was less than 1\,\% over a period of 12\,h.
The average polarization was $P_T=0.80\pm 0.04$.

The deflection of the incoming muons caused by the transverse
dipole field was 2.25~mrad, and was compensated by an additional magnet,
 installed
7\,m upstream of the target. The reconstruction software was modified to
account for the curvature of the beam tracks
in the region between the two sets of scintillator hodoscopes used
to determine their direction.
Track reconstruction and vertex fitting inside the dipole field  were tested
by a Monte~Carlo simulation, and were found to perform just as well as
in the case of the solenoid field used for longitudinal asymmetry measurements.

The incident muons mostly come from pion decay and
are naturally polarized in the longitudinal
direction. The average beam polarization at 100 GeV was determined from the
positron energy spectrum in the decay $\mu^+ {\rightarrow}\, e^+ \nu_e
 \overline{\nu}_\mu $, and
found to be $P_B$~=~$-0.82~\pm~0.06$ \cite{SMC93,SMC94A}, in good agreement
with the Monte Carlo simulation of the beam transport~\cite{LG} .

\begin{figure}[ht]
 \begin{center}
 \mbox{ \epsfig{file=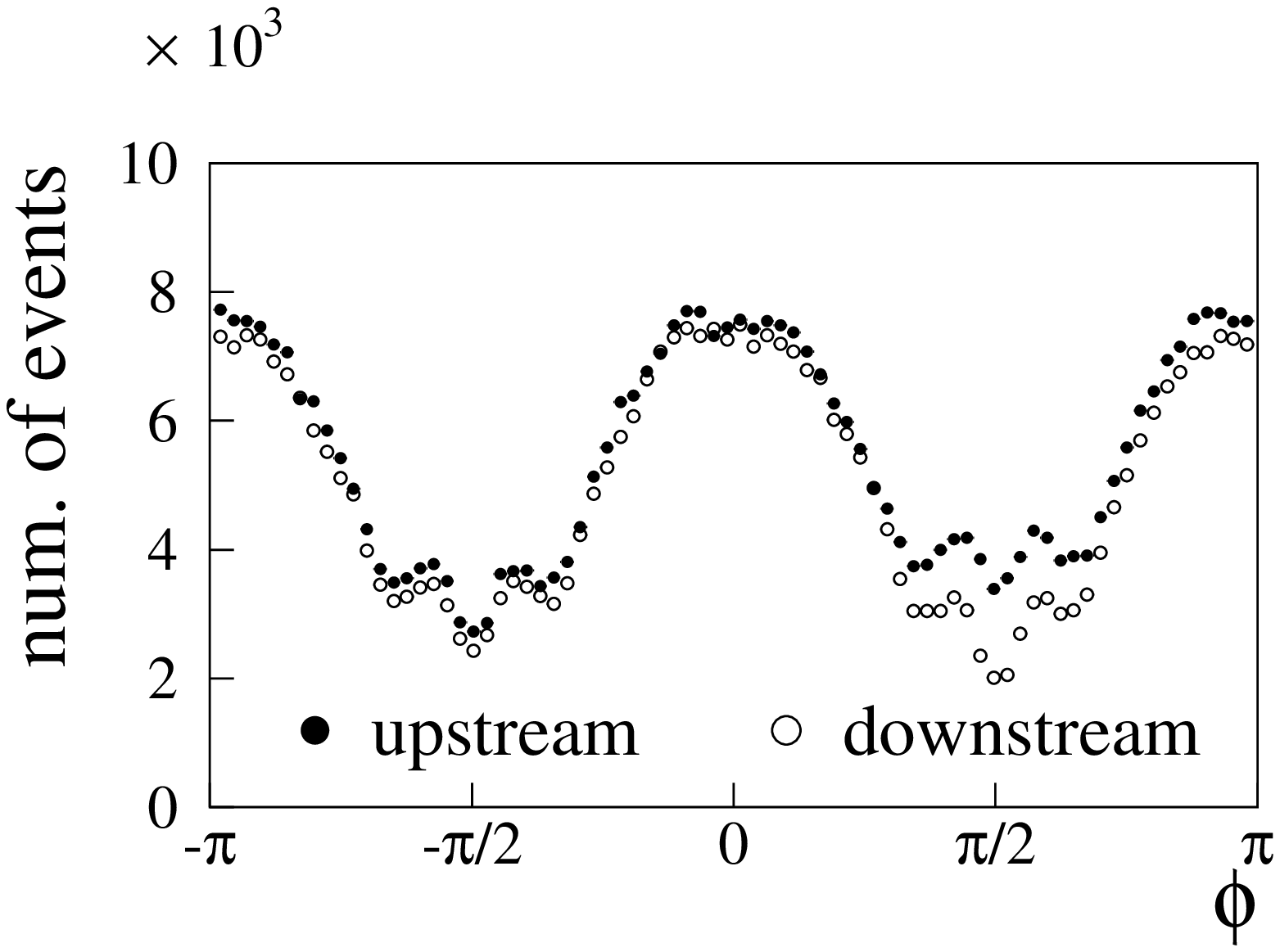,height=8cm,width=12cm}}
 \end{center}
\caption{
    The event distribution as a function of the azimuthal angle $\phi$,
    for interactions in the upstream and downstream targets.
    The value $\phi=0$ corresponds to a scattered muon with transverse
    momentum parallel to the target polarization.
    The observed distribution
    mainly reflects the trigger acceptance.}
\label{fig1}
\end{figure}

The events were required to
satisfy $y<0.9$,~$E_{\mu'}>15$~GeV, and $\nu>10$~GeV,
in order to avoid large radiative corrections,
to eliminate muons originating from the decay of pions
produced in the target,
and events with poor kinematic resolution.
A total of $8.7\times 10^5$ events was obtained in the range
$0.006<x<0.6$ and $1<Q^2<30$~GeV$^2$.
The event distribution as a function of the angle $\phi$ is shown in
Fig.~\ref{fig1}.
Most events are collected in the regions where the angle $\phi$ is
close to zero or $\pi$ because the trigger conditions predominantly
select muons scattered in a direction close to the vertical plane.
This optimizes our measurement, because only muons scattered in a plane
close to the polarization plane contribute effectively to the
asymmetry.

%% Transverse cross section asymmetries are more difficult to measure than
%% longitudinal ones.
%% The kinematic factors $\sqrt{1-y}/(1-y/2)$ in Eq.~(\ref{d})
%% and $\gamma$ in Eq.~(\ref{A2}),
%% reduce the magnitude of the measured $A_\perp$ at low $x$ or at
%% large $Q^2$, see Eq.~(\ref{AG2}). In addition,
%\newpage
The asymmetries were obtained separately for each target cell.
The raw asymmetry $A_m$ is measured in bins of $x, Q^2$, and $\phi$ from
the number of scattered muons
($N_1,N_2$) in the azimuthal directions ($\phi$,$\pi-\phi$), and the
corresponding counts ($N_1',N_2'$)
evaluated after a reversal of the polarization. Assuming that the ratio of the
spectrometer acceptances at angles $\phi$ and $\pi-\phi$ is the same during
the periods before and after polarization
reversal, we obtain

\begin{equation}
\sqrt {\frac{N_1 N_2'}{N_2 N_1'}} =
\frac{1+A_m(x,Q^2,\phi)}{1-A_m(x,Q^2,\phi)}\
\label{araw}
\end{equation}

\noindent
The transverse asymmetry is then
\begin{equation}
    A_{\perp}(x,Q^2) =
        \frac{1}{\cos\phi} \frac{A_m(x,Q^2,\phi)}{P_B |P_T| f}\;,
\label{at_araw}
\end{equation}
where $P_B$ and $P_T$ are the beam and target polarizations, and $f$ is
the fraction of polarizable protons in the target material~$(f\cong0.12)$.
The values of $A_\perp$ corresponding to the same $x$ and $Q^2$ are averaged
 over the $\phi$
intervals, and over the five subsamples defined by the ten
polarization reversals.

The asymmetry $A_2$ is extracted from $A_\perp$ and $A_\parallel$ in
 ($x$,$Q^2$)
 bins, using Eqs.~(\ref{AG1}) and (\ref{AG2}).  For $A_\parallel$ we use
previous measurements on longitudinally polarized targets~\cite{SY,AS88,SMC94}.
These experiments have published $A_1$, extracted through the relation
$A_1=A_\parallel /D $, where the $A_2$ contribution is neglected. We
 parametrize
 the $A_1$ data and recover $A_\parallel$
through the same relation.

Since no $Q^2$ dependence is observed within the errors, we present the average
 of $A_2$ in each bin of $x$ (see Table~\ref{a2numbs}).
If the $\overline{g_2}$ contribution can be neglected in Eq.~(\ref{g2}) and
$A_1$ is independent of $Q^2$, $ \sqrt{Q^2} A_2$  is expected to scale.
 However,
 if we average $ \sqrt{Q^2} A_2$, instead of $A_2$, we obtain the same results.

The dominant systematic error on $A_\perp$ is the variation
of the ratio of acceptances in the upper and lower parts of
the spectrometer between polarization
reversals. The resulting false asymmetries have been studied
using real data and a Monte~Carlo simulation, and found to be negligible
compared to the
statistical error.
Radial effects and overall variations in detector efficiencies
do not cause false asymmetries because they affect
the upper and lower part of the spectrometer in the same
way. The radiative corrections to $A_{\perp}$ were calculated with the method
of Ref.~\cite{SHUM} and found to be smaller than 0.001.

\begin{figure}[ht]
 \begin{center}
 \mbox{ \epsfig{file=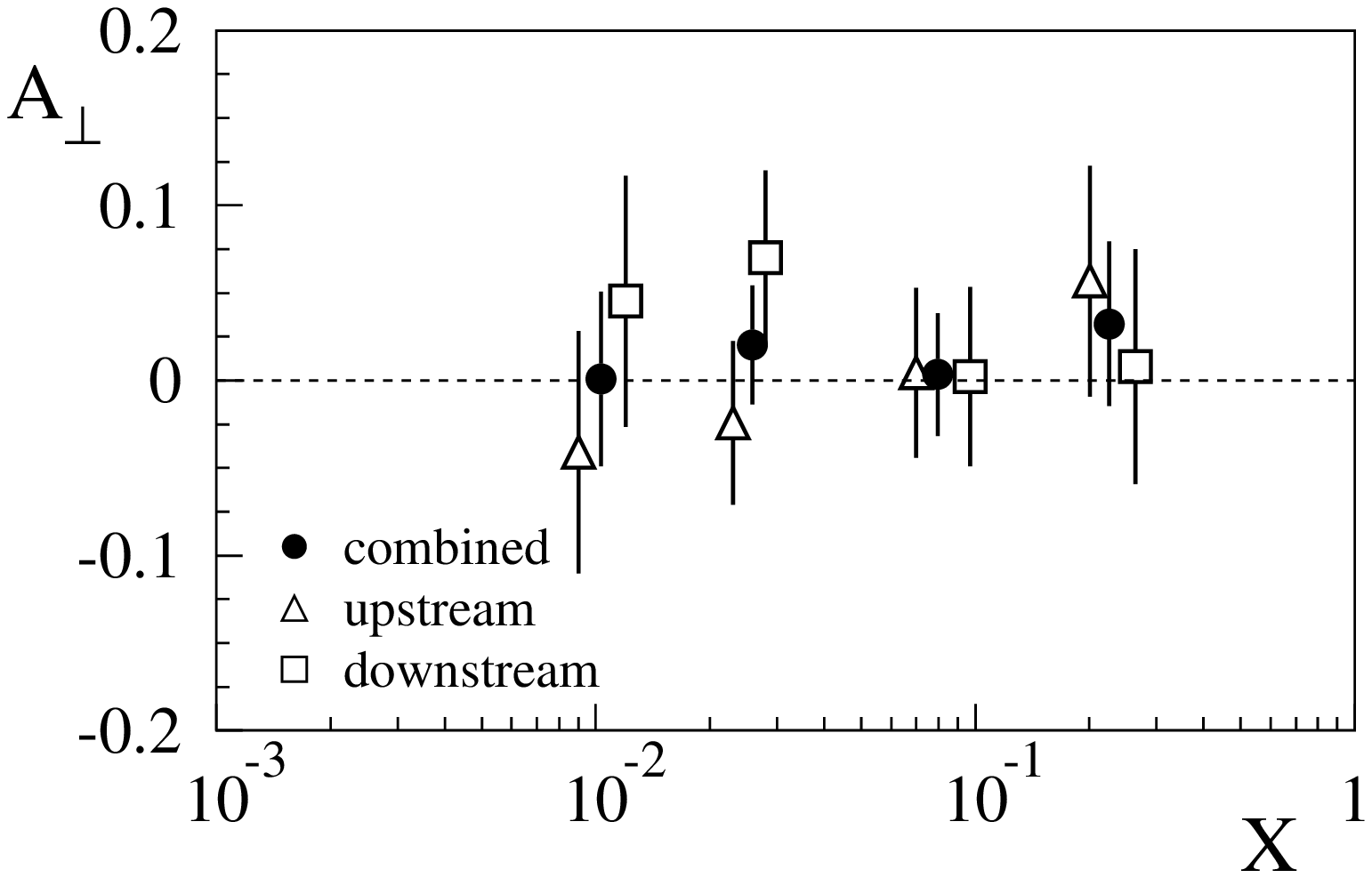,height=8cm,width=12cm}}
 \end{center}
    \caption{
    The transverse cross-section asymmetry
    $A_{\perp}$ as a function of $x$ for interactions in the
    upstream and downstream targets and for the combined data.
    The error bars represent the statistical errors.}
\label{fig2}
\end{figure}

The resulting values for $A_\perp$ are shown in Fig.~\ref{fig2}, for four
intervals of $x$.
They are consistent for the two parts of the target and are compatible with
 zero.

\begin{figure}[hbt]
 \begin{center}
 \mbox{ \epsfig{file=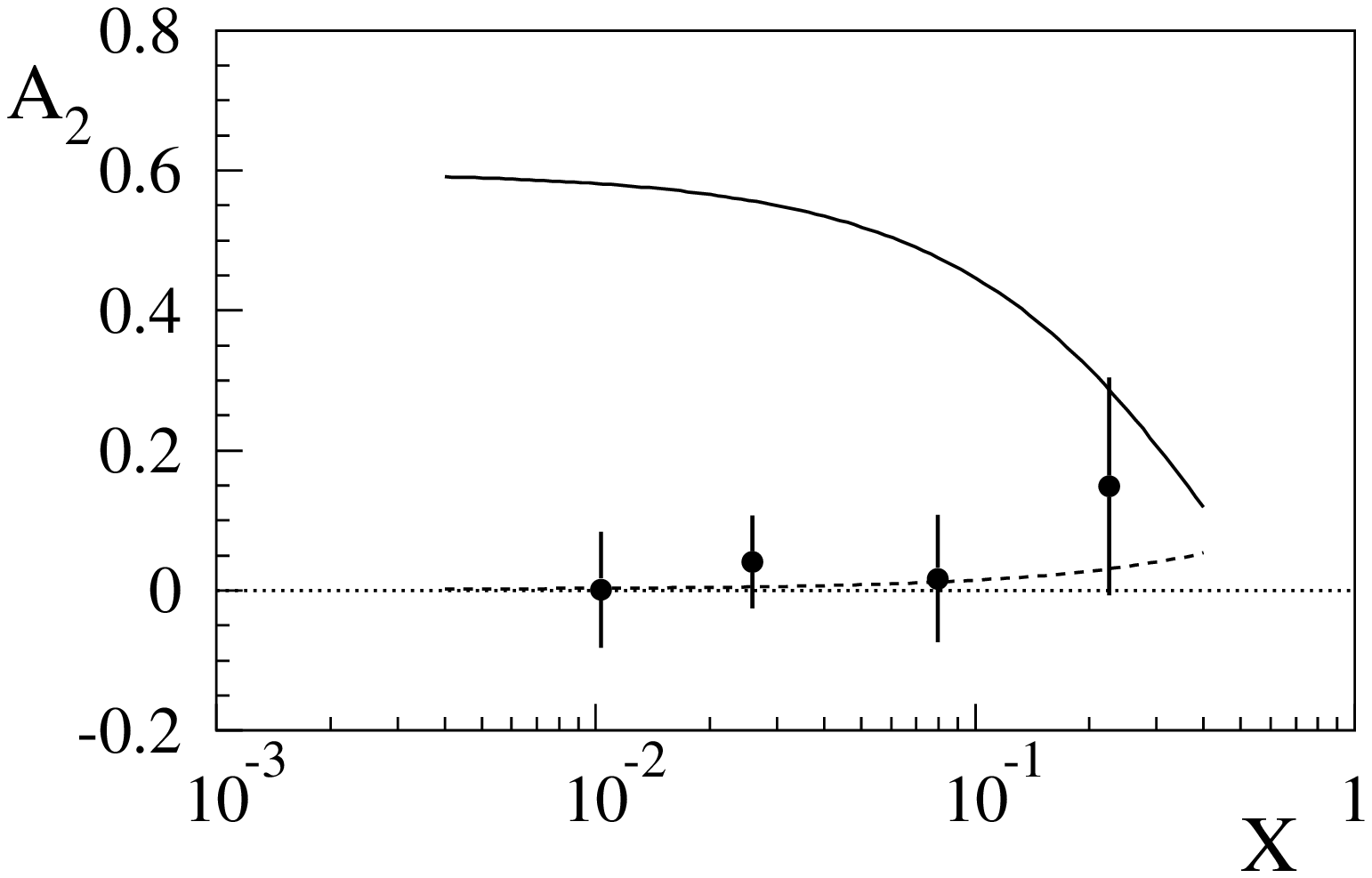,height=8cm,width=12cm}}
 \end{center}
    \caption[]{
    The asymmetry $A_2$ as a function of $x$. The solid line shows
    $\sqrt{R}$ from the SLAC parametrization~[12],
    and the dashed line the results obtained with
    $\overline{g_2}=0$ in Eq.~(1).
    The error bars represent statistical errors, only.}
\label{fig3}
\end{figure}

The corresponding values of $A_2$ are presented in Fig.~\ref{fig3} and
 Table~\ref{a2numbs}.
The limit imposed by the positivity condition (Eq.~\ref{UPPER})
is also shown in Fig.~\ref{fig3}.
The present measurements constrain the
asymmetry function  $A_2$ to values much smaller than the
positivity limit. At 90~\% confidence level, we obtain $A_2<0.16$
for $x<0.15$, and $A_2<0.4$ for $x>0.15$. These improved limits
have been used in Ref. \cite{SMC94}
for the evaluation of $g_1$.
The expected
values of $A_2$ obtained by considering only the first
term of $g_2$ in Eq.~\ref{g2}
are also shown in Fig.~\ref{fig3}, and are in good agreement with the data.
They have been calculated from the values of
$g_1$ from Refs.~\cite{SY,AS88,SMC94} assuming that $g_1$ scales with $F_1$.
Additional sources of systematic errors
in $A_2$ are the parametrization of longitudinal asymmetries, and the
 uncertainty
in $R$. Both effects are smaller than 10~\% of the statistical error.

The values obtained for $g_2$ are listed in
Table~\ref{a2numbs}, together with those of $g_{2}^{\rm ww}$ (Eq.~\ref{g2ww})
evaluated at the same $x$ and $Q^2$.
The two sets are compatible within the statistical errors. Therefore
the present data do not show evidence for
$\overline{g_2}$ to be different from zero.
The limited accuracy of the present data does not allow sensitive
tests of specific predictions, such as the partial cancellation between
the two terms of $g_2$ predicted by the bag models of
Refs.~\cite{STRAT,JAJI}.
Large contributions of $\overline{g_2}$ close to the positivity
limit are excluded in the $x$~range covered.
However, $\overline{g_2}$ about one order of
magnitude larger than  $g_{2}^{\rm ww}$ would still be allowed within
the statistical errors.

When the minimum $Q^2$ requirement
is reduced to 0.5~GeV$^2$,
the data with $Q^2 < 1$~GeV$^2$ in the lowest $x$-bin yield  $A_2=0.05 \pm
 0.10$, which is consistent with the value for $Q^2 > 1$~GeV$^2$.
If we neglect the $Q^2$ dependence within the kinematic range of the present
 data,
the results of Table~\ref{a2numbs} can be used to calculate the limits for the
 integral of $g_2$,
\begin{equation}
-0.9 < \int_{0.006}^{0.6} g_2 dx < 1.9\;,
\label{int_r1}
\end{equation}
at 90 \% confidence level.
We do not attempt the evaluation of the first moment of $g_2$, in order
to test the Burkhardt--Cottingham sum rule, because we are not aware of
theoretical predictions for the behaviour of $g_2$ as $x \rightarrow 0$.

In summary, we have presented the first measurement of transverse
asymmetries in deep inelastic lepton--proton scattering. The
virtual-photon asymmetry $A_2$ is found to be significantly smaller
than its positivity limit $\sqrt{R}$. This result reduces the
uncertainty in the determination of the structure function $g_1$ from
longitudinal polarization measurements. In addition, bounds are
obtained for the spin-dependent structure function
$g_2$, which exclude large twist-3 contributions.

\begin{table}
\caption[.]{Results on the spin asymmetry $A_2$
and the structure functions $g_2$ and $g_{2}^{\rm ww}$, where
$g_{2}^{\rm ww}$ has been calculated from the results of
Ref.~\cite{SY,AS88,SMC94}.  Only the statistical errors are given.}
\begin{center}
\begin{tabular}{|c|c|c|c|c|c|}    \hline
$x$ interval & $\langle x \rangle$ & $\langle Q^2($GeV$^2)\rangle$ & $A_2$ &
 $g_{2}$ & $g_{2}^{\rm ww}$  \\
\hline
0.006 -- 0.015 & 0.010 & ~1.4 & $0.002\pm 0.083$ & $1.2\pm 61~$ &~~$0.73\pm
0.10$
 \\
0.015 -- 0.050 & 0.026 & ~2.7 & $0.041\pm 0.066$ & $7.0\pm 12~$ &~~$0.47\pm
0.09$
 \\
0.050 -- 0.150 & 0.080 & ~5.8 & $0.017\pm 0.091$ & $0.2\pm 2.9$ &~~$0.15\pm
 0.02$
 \\
0.150 -- 0.600 & 0.226 & 11.8 & $0.149\pm 0.156$ & $0.5\pm 0.8$ & $-0.10\pm
 0.02$ \\
\hline
\end{tabular}
\end{center}
\label{a2numbs}
\end{table}

\end{document}